\documentclass[english]{article}
\usepackage[T1]{fontenc}
\usepackage[latin9]{inputenc}
\usepackage{amstext}
\usepackage{amssymb}
\usepackage{graphicx}
\usepackage{esint}
\usepackage{tikz}
\usepackage{braket}
\makeatletter
\usepackage{amsfonts,amssymb}
\usepackage{latexsym}
\usepackage{epsfig}
\usepackage{cite}
\usepackage{graphicx}
\usepackage[colorlinks,linkcolor=blue]{hyperref}
\newcommand{\be}{\begin{equation}}
\newcommand{\ee}{\end{equation}}

\makeatother

\usepackage{babel}
\begin{document}
{}~ \hfill\vbox{\hbox{CTP-SCU/2023009}}
\break \vskip 3.0cm 
\centerline{\Large \bf  Holographic $T\bar{T}$ deformed entanglement entropy  in $\text{dS}_3/\text{CFT}_2$   }
\vspace*{10.0ex} 
\centerline{\large Deyou Chen$^{a,b,*}$ Xin Jiang$^{c,\dagger}$ and Haitang Yang$^{c,\ddagger}$} 
\vspace*{7.0ex} \vspace*{4.0ex} 

\centerline{\large \it $^a$School of Science, Xihua University, Chengdu 610039, China} 
\vspace*{1.0ex}
\centerline{\large \it $^b$Key Laboratory of High Performance Scientific Computation,}
\vspace*{1.0ex}
\centerline{\large \it Xihua University, Chengdu 610039, China}
\vspace*{1.0ex}
\centerline{\large \it $^c$College of Physics, Sichuan University, Chengdu, 610065, China} 

\vspace*{1.0ex} \vspace*{4.0ex} 

\centerline{$^*$deyouchen@hotmail.com, $^\dagger$domoki@stu.scu.edu.cn, $^\ddagger$hyanga@scu.edu.cn} 
\vspace*{10.0ex} \centerline{\bf Abstract} \bigskip \smallskip 

In this paper, based on the $T\bar{T}$ deformed version of $\text{dS}_3/\text{CFT}_2$ correspondence,  we calculate the pseudoentropy for an entangling surface consisting of two antipodal points on a sphere and find it is exactly dual to the complex geodesic in the bulk.

\vfill 
\eject
\baselineskip=16pt


{}

{}

\section{Introduction}

\label{sec:Introduction}

The study of quantum gravity in de Sitter space has generated much
interest in recent years, particularly due to its potential relevance
for inflationary cosmology and cosmic acceleration. One promising
method to comprehend de Sitter space is through the dS/CFT
 correspondence \cite{Strominger:2001pn}. It
is a conjectured equivalence between a gravitational theory in de
Sitter space and a conformal field theory residing on its boundary.
The dS/CFT correspondence is a generalization of the well-known AdS/CFT correspondence\cite{Maldacena:1997re,Witten:1998qj},
which has been extensively studied in string theory and provided numerous
insights into extracting the nature of quantum gravity from its dual
CFT. However, the dS/CFT correspondence is not as well understood
as the AdS/CFT correspondence, as there are only limited explicit
examples of CFTs that are dual to de Sitter spacetime. Recently, a
remarkable and explicit example has been constructed for the $\text{dS}_{3}/\text{CFT}_{2}$
correspondence \cite{Hikida:2021ese,Hikida:2022ltr}, where the dual
CFT resides on the past/future boundary of de Sitter spacetime.

Starting with the three-dimensional de Sitter spacetime, we 
take a static compact slice $\Sigma_{t}$ of constant time.
Clearly, each $\Sigma_{t}$ has a Riemannian metric $\gamma$ and
a second fundamental form $K$. In the canonical formalism for gravity,
the quantum state residing on $\Sigma_{t}$ can be described by the
Hartle-Hawking wavefunction $\Psi_{\text{dS}}\left[\gamma\right]$.
Following \cite{Hikida:2021ese,Hikida:2022ltr}, and neglecting the
contributions of bulk matter fields, we could obtain a calculable
example of $\text{dS}_{3}$/$\text{CFT}_{2}$ correspondence described
by
\begin{equation}
\Psi_{\text{dS}}\left[\gamma\right]=Z_{\text{CFT}}\left[\gamma\right],\quad t\rightarrow\infty
\end{equation}
where $Z_{\text{CFT}}$ is the partition function of the dual $\text{CFT}_{2}$
living on $\Sigma_{\infty}$. 

\begin{figure}
\includegraphics[width=1\textwidth]{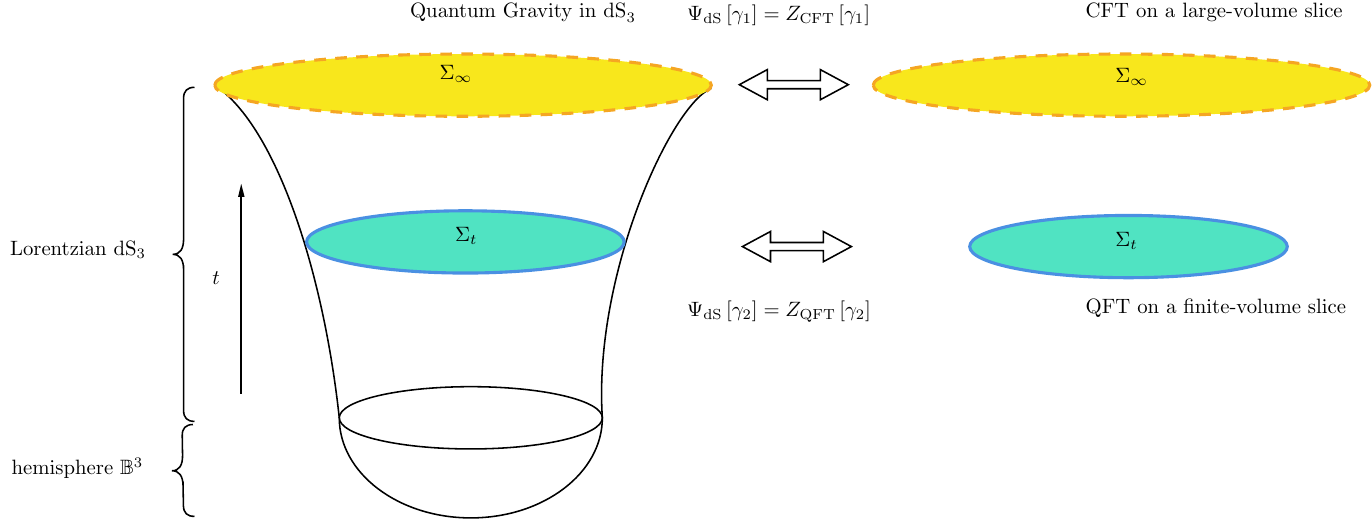}\label{img1}

\caption{The $T\bar{T}$-deformed version of the $\text{dS}_{3}$/$\text{CFT}_{2}$
correspondence, with the Lorentzian time $t$ in the global coordinates
of the $\text{dS}_{3}$ spacetime.}
\end{figure}

In this paper, we aim to further explore the scenario described above.
Typically, it is not necessary to confine the slice $\Sigma_{t}$ at
the future infinity, which leads to a natural extension of the $\text{dS}_{3}$/$\text{CFT}_{2}$
correspondence:
\begin{equation}
\Psi_{\text{dS}}\left[\gamma\right]=Z_{\text{QFT}}\left[\gamma\right],\label{eq:ttbar_dS}
\end{equation}
where $Z_{\text{QFT}}$ is the partition function of the dual quantum
field theory(QFT) living on a finite-volume slice $\Sigma_{t}$. 
The
dual QFT could be defined as a $2d$ CFT deformed by the $T\bar{T}$
operator \cite{Zamolodchikov:2004ce,Cavaglia:2016oda,Smirnov:2016lqw}
that generates a trajectory in the space of field theory,
\begin{equation}
\frac{\partial}{\partial\lambda}\log Z\left(\lambda\right)=-2\pi\int_{\Sigma}d^{2}x\sqrt{\gamma}\left\langle T\bar{T}\right\rangle _{\lambda}.\label{eq:tt}
\end{equation}
At the first order of the deformation parameter $\lambda$,
the deformed theory, perturbatively, could be written as
\begin{equation}
\log Z\left(\lambda\right)=\log Z\left(\lambda=0\right)-2\pi\lambda\int_{\Sigma}d^{2}x\sqrt{\gamma}\left\langle T\bar{T}\right\rangle _{\lambda=0} + {\mathcal O}(\lambda^2),\label{eq:per-tt}
\end{equation}
where $\left\langle T\bar{T}\right\rangle _{\lambda=0}$ is defined
by the stress tensor of the undeformed theory as $\left\langle T\bar{T}\right\rangle _{\lambda=0}\equiv\left\langle T\bar{T}\right\rangle =\frac{1}{8}\left[\left\langle T^{ab}\right\rangle \left\langle T_{ab}\right\rangle -\left\langle T_{a}^{a}\right\rangle ^{2}\right]$.
In recent years, the $T\bar{T}$ deformation has been widely studied
\cite{Dubovsky:2017cnj,Donnelly:2018bef,Kraus:2018xrn,Aharony:2018vux,Conti:2018tca,Cardy:2018sdv,Conti:2018jho,Bonelli:2018kik,Aharony:2018bad,Datta:2018thy,Chen:2019mis,Conti:2019dxg,Guica:2019nzm,Ishii:2019uwk,Grieninger:2019zts,Jeong:2019ylz,Jiang:2019epa,He:2019vzf,Jiang:2019tcq,Pozsgay:2019ekd,Grado-White:2020wlb,Khoeini-Moghaddam:2020ymm,Caputa:2020lpa,Medenjak:2020ppv,Li:2020pwa,He:2020qcs,Ceschin:2020jto,Jiang:2020nnb,He:2022xkh,Cardona:2022cmh,Aramini:2022wbn,Hou:2023ytl,Jiang:2023ffu,He:2023eap,He:2023wko,Castro-Alvaredo:2023jbg,Tian:2023fgf},
due to its integrability and its applications in holography. Our proposal
is a natural extension of the cutoff-AdS/$T\bar{T}$-deformed-CFT
correspondence \cite{McGough:2016lol} to de Sitter spacetime. Additionally,
the $T\bar{T}$-deformed version (\ref{eq:ttbar_dS}) of $\text{dS}_{3}$/$\text{CFT}_{2}$
is remarkably coincident with the Cauchy slice holography\cite{Araujo-Regado:2022gvw,Araujo-Regado:2022jpj},
where time serves as the emergent direction. The $T\bar{T}$-deformed
version (\ref{eq:ttbar_dS}) of $\text{dS}_{3}$/$\text{CFT}_{2}$
is illustrated in Figure \ref{img1}. Note that the deformation parameter $\lambda$ is on the order of $\mathcal{O}(1/c)$, and in the limit of large $c$, the deformed theory is simply defined by equation (\ref{eq:per-tt}). In this scenario, the $T\bar{T}$ flow has been demonstrated to non-perturbatively match with bulk computations of dS$_3$ with a finite temporal cut-off, analogous to the situation in AdS/CFT \cite{McGough:2016lol}. This justifies our choice of the time $t$ on the hypersurface $\Sigma_t$ to be finite. However, beyond the large $c$ limit, one must define the $T\bar{T}$ deformed theory using equation (\ref{eq:tt}), which generically cannot be solved completely.

It is clear that the $T\bar{T}$ deformation is irrelevant in the
renormalization group sense. This implies that the $T\bar{T}$ deformation
leads no consequence in IR but does affect UV physics. Among the various
deformable physical quantities in the UV region, a particularly
important one is the entanglement entropy. In the dS/CFT correspondence,
the dual CFTs turn to be non-unitary \cite{Strominger:2001pn,Maldacena:2002vr,Hikida:2021ese,Hikida:2022ltr}.
To characterize the degrees of freedom  in a non-unitary
CFT, complex-valued entanglement entropies, namely pseudoentropy
\cite{Nakata:2020luh,Mollabashi:2020yie,Nishioka:2021cxe,Mollabashi:2021xsd,Narayan:2022afv,Doi:2022iyj,
Doi:2023zaf,Narayan:2023ebn,Jiang:2023ffu,Kawamoto:2023nki,He:2023wko,He:2023eap,Chu:2023zah},
are needed. In other words, the pseudoentropy can be viewed as a well-defined
entanglement entropy in the $T\bar{T}$-deformed version of the dS$_{3}$/CFT$_{2}$
correspondence. It is then interesting to explore how the holographic
entanglement entropy \cite{Ryu:2006bv,Ryu:2006ef} behaves in the $T\bar{T}$-deformed
version of the dS$_{3}$/CFT$_{2}$. In this paper, our main goal
is to calculate the entanglement entropy in the $T\bar{T}$-deformed
field theory and compare it with geodesics in the dS$_{3}$ bulk. 

In section $2$, we give a brief review of the $T\bar{T}$  deformed version of dS/CFT. 
In section \ref{sec:Holographic-Entanglement-Entropy}, we calculate
the pseudoentropy for an entangling surface consisting of two antipodal
points on a sphere $\mathbb{S}^{2}$, and we find that the entanglement
entropy does perfectly match the length of the complex geodesic connecting
these antipodal points in dS$_{3}$.

\section{Wheeler-DeWitt equation and $T\bar{T}$ flow}


\label{sec:Wheeler-DeWitt-equation-=000026}

We first briefly review the $T\bar{T}$  deformed version of dS/CFT in this section.
In the canonical formalism of three-dimensional pure gravity, the
Hartle-Hawking wavefunction $\widetilde{\Psi}\left[\gamma\right]$
should obey the Wheeler-DeWitt equation
\begin{equation}
\mathcal{H}\widetilde{\Psi}\left[\gamma\right]=\left\{ \frac{16\pi G_{N}}{\sqrt{\gamma}}\left(\Pi^{ab}\Pi_{ab}-\Pi_{a}^{a}\Pi_{b}^{b}\right)-\frac{\sqrt{\gamma}}{16\pi G_{N}}\left(\mathcal{R}\left[\gamma\right]-2\Lambda\right)\right\} \widetilde{\Psi}\left[\gamma\right]=0,
\end{equation}
where $\mathcal{R}\left[\gamma\right]$ is the Ricci scalar, $\Lambda=\ell_{\text{dS}}^{-2}$
is the cosmological constant of the de Sitter spacetime, and $\Pi^{ab}$
is the momentum conjugate to the metric $\gamma_{ab}$:
\begin{equation}
\Pi^{ab}=-\text{i}\frac{\delta}{\delta\gamma_{ab}}=\frac{\sqrt{\gamma}}{16\pi G_{N}}\left(K^{ab}-K_{c}^{c}\gamma^{ab}\right).
\end{equation}
 The standard quasilocal stress tensor can be defined as:
\begin{equation}
T^{ab}=-\frac{2}{\sqrt{\gamma}}\frac{\delta}{\delta\gamma_{ab}}=-\frac{2\text{i}}{\sqrt{\gamma}}\Pi^{ab},
\end{equation}
which coincides with the field-theoretic definition. To require
the finiteness of the quasilocal stress tensor at the future infinity
$\Sigma_{\infty}$, one needs to perform a canonical transformation
\cite{Araujo-Regado:2022jpj,Witten:2022xxp},
\begin{equation}
\widetilde{\Psi}=\exp\left(-\frac{\text{i}}{8\pi G_{N}\ell_{\text{dS}}}\int_{\Sigma}d^{2}x\sqrt{\gamma}\right)\Psi,\label{eq:cano-trans}
\end{equation}
which leads a shift on the momentum
\begin{equation}
\Pi^{ab}\rightarrow\Pi^{ab}+\frac{\sqrt{\gamma}}{16\pi G_{N}\ell_{\text{dS}}}\gamma^{ab}.
\end{equation}
Therefore, the Wheeler-DeWitt equation could be rewritten as 
\begin{equation}
\left\{ -\frac{2}{\sqrt{\gamma}}\Pi_{a}^{a}+\frac{16\pi G_{N}\ell_{\text{dS}}}{\text{det}\gamma}\left(\Pi^{ab}\Pi_{ab}-\Pi_{a}^{a}\Pi_{b}^{b}\right)-\frac{\ell_{\text{dS}}}{16\pi G_{N}}\mathcal{R}\left[\gamma\right]\right\} \Psi\left[\gamma\right]=0.
\end{equation}
By using the quasilocal stress tensor, the equation is simply
\begin{equation}
T_{a}^{a}=\frac{\text{i}\,\ell_{\text{dS}}}{16\pi G_{N}}\mathcal{R}\left[\gamma\right]+\text{i}\,4\pi G_{N}\ell_{\text{dS}}\left(T^{ab}T_{ab}-T_{a}^{a}T_{b}^{b}\right).\label{eq:flow}
\end{equation}

On the other hand, in the $T\bar{T}$ deformed field theory,  when the deformation parameter $\lambda$
is small\footnote{Our approach is limited to perturbation theory because a nonperturbative completion of the $T\bar{T}$  deformation (\ref{eq:tt}) is unknown. In our analysis, we operate under the assumption that the Zamolodchikov's factorization formula remains valid in the context of dS/CFT, particularly when considering the large $c$ limit.}, one can
rewrite eqn.(\ref{eq:tt}) as
\begin{equation}
\log Z_{\text{QFT}}=\log Z_{\text{CFT}}-2\pi\lambda\int_{\Sigma}d^{2}x\sqrt{\gamma}\left\langle T\bar{T}\right\rangle .
\end{equation}
By the definition of the trace of the stress tensor
\begin{equation}
T_{a}^{a}=-2\frac{\gamma_{ab}}{\sqrt{\gamma}}\frac{\delta}{\delta\gamma_{ab}}\log Z,
\end{equation}
and the famous Weyl anomaly
\begin{equation}
\left(T_{a}^{a}\right)_{\text{CFT}}=-\frac{c}{24\pi}\mathcal{R}\left[\gamma\right],
\end{equation}
the trace flow equation for deformed theory is 
\begin{equation}
\left\langle T_{a}^{a}\right\rangle =-\frac{c}{24\pi}\mathcal{R}\left[\gamma\right]-\frac{\pi\lambda}{2}\left(\left\langle T^{ab}\right\rangle \left\langle T_{ab}\right\rangle -\left\langle T_{a}^{a}\right\rangle \left\langle T_{b}^{b}\right\rangle \right).
\label{eq:traceflow}
\end{equation}
Here, all stress tensors emanate from the deformed theory residing on a two-sphere. This alignment is in accordance with the capability of $T\bar{T}$ deformation to be defined on compact backgrounds \cite{Cavaglia:2016oda,Donnelly:2018bef,He:2022xkh}.
Relating to eqn.(\ref{eq:flow}), we immediately find the identifications
between field-theoretic quantities and gravitational quantities
\begin{equation}
c=-\text{i}\frac{3\ell_{\text{dS}}}{2G_{N}}=-\text{i}\,c_{\text{dS}},\quad\lambda=-\text{i}\,8G_{N}\ell_{\text{dS}}=-\text{i}\,\lambda_{\text{dS}},\label{eq:relation}
\end{equation}
 where the Brown-Henneaux central charge \cite{Brown1986} turns to
be imaginary-valued in the de Sitter context \cite{Maldacena:2002vr},
and the deformation parameter is also imaginary-valued. The deformation parameter $\lambda$ remains unrelated to the time variable $t$ due to our selection of the seed CFT residing on the hypersurface $\Sigma_t$, as opposed to $\Sigma_\infty$. It is noteworthy that these distinct choices of the seed CFT are connected through a straightforward Weyl rescaling on the hypersurface. Furthermore,
the momentum constraint for the Hartle-Hawking wavefunction,
\begin{equation}
\mathcal{D}^{a}\Psi\left[\gamma\right]=-2\left(\nabla_{b}\Pi^{ab}\right)\Psi\left[\gamma\right]=0,
\end{equation}
 can be easily interpreted as the conversation law of the stress tensor
in field theory
\begin{equation}
\nabla_{b}\left\langle T^{ab}\right\rangle =0.
\end{equation}
The wavefunction  $\Psi\left[\gamma\right]$ should be invariant under diffeomorphisms of $\Sigma_t$, given that $\mathcal{D}^{a}$ serves as the generator of diffeomorphisms. In simpler terms,  $\Psi\left[\gamma\right]$  is a function on the space of metrics modulo diffeomorphisms.
Even though the dual field theory is non-unitary, the dynamical inner
product $\left\langle \Psi|\Psi\right\rangle $ is Hermitian and positive-semidefinite,
which indicates that we still have the bulk unitarity \cite{Araujo-Regado:2022gvw,Araujo-Regado:2022jpj}.

\section{Holographic Entanglement Entropy}

\label{sec:Holographic-Entanglement-Entropy}

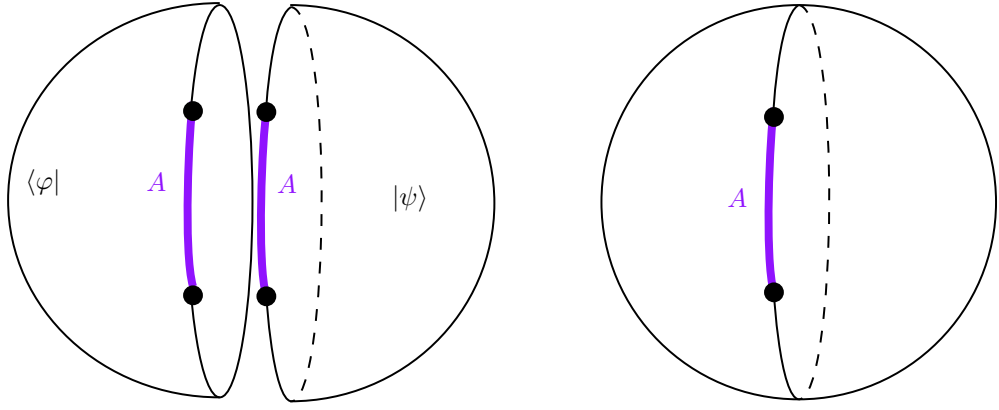
\begin{figure}[h]

\tikzset{every picture/.style={line width=0.75pt}} 
\begin{center}
\begin{tikzpicture}[x=0.75pt,y=0.75pt,yscale=-1,xscale=1]

\draw    (177.5,270) .. controls (160,268) and (157,82) .. (177.5,71) ;
\draw [color={rgb, 255:red, 144; green, 19; blue, 254 }  ,draw opacity=1 ][line width=3]    (165,124) .. controls (163,136) and (160,200) .. (165,217) ;
\draw  [dash pattern={on 4.5pt off 4.5pt}]  (177.5,71) .. controls (199,74) and (197,270) .. (177.5,270) ;
\draw  [draw opacity=0] (141.5,268) .. controls (141.5,268) and (141.5,268) .. (141.5,268) .. controls (141.5,268) and (141.5,268) .. (141.5,268) .. controls (82.54,268) and (34.75,223.45) .. (34.75,168.5) .. controls (34.75,113.55) and (82.54,69) .. (141.5,69) -- (141.5,168.5) -- cycle ; \draw   (141.5,268) .. controls (141.5,268) and (141.5,268) .. (141.5,268) .. controls (141.5,268) and (141.5,268) .. (141.5,268) .. controls (82.54,268) and (34.75,223.45) .. (34.75,168.5) .. controls (34.75,113.55) and (82.54,69) .. (141.5,69) ;  
\draw   (125,169) .. controls (125,114.32) and (132.39,70) .. (141.5,70) .. controls (150.61,70) and (158,114.32) .. (158,169) .. controls (158,223.68) and (150.61,268) .. (141.5,268) .. controls (132.39,268) and (125,223.68) .. (125,169) -- cycle ;
\draw  [draw opacity=0] (177.03,70) .. controls (177.03,70) and (177.03,70) .. (177.03,70) .. controls (233.9,70) and (280,114.77) .. (280,170) .. controls (280,225.23) and (233.9,270) .. (177.03,270) -- (177.03,170) -- cycle ; \draw   (177.03,70) .. controls (177.03,70) and (177.03,70) .. (177.03,70) .. controls (233.9,70) and (280,114.77) .. (280,170) .. controls (280,225.23) and (233.9,270) .. (177.03,270) ;  
\draw [color={rgb, 255:red, 144; green, 19; blue, 254 }  ,draw opacity=1 ][line width=3]    (128,119) .. controls (126,131) and (123,195) .. (128,212) ;
\draw  [fill={rgb, 255:red, 0; green, 0; blue, 0 }  ,fill opacity=1 ] (123.5,216.5) .. controls (123.5,214.01) and (125.51,212) .. (128,212) .. controls (130.49,212) and (132.5,214.01) .. (132.5,216.5) .. controls (132.5,218.99) and (130.49,221) .. (128,221) .. controls (125.51,221) and (123.5,218.99) .. (123.5,216.5) -- cycle ;
\draw  [fill={rgb, 255:red, 0; green, 0; blue, 0 }  ,fill opacity=1 ] (123.5,123.5) .. controls (123.5,121.01) and (125.51,119) .. (128,119) .. controls (130.49,119) and (132.5,121.01) .. (132.5,123.5) .. controls (132.5,125.99) and (130.49,128) .. (128,128) .. controls (125.51,128) and (123.5,125.99) .. (123.5,123.5) -- cycle ;
\draw  [fill={rgb, 255:red, 0; green, 0; blue, 0 }  ,fill opacity=1 ] (160.5,217) .. controls (160.5,214.51) and (162.51,212.5) .. (165,212.5) .. controls (167.49,212.5) and (169.5,214.51) .. (169.5,217) .. controls (169.5,219.49) and (167.49,221.5) .. (165,221.5) .. controls (162.51,221.5) and (160.5,219.49) .. (160.5,217) -- cycle ;
\draw  [fill={rgb, 255:red, 0; green, 0; blue, 0 }  ,fill opacity=1 ] (160.5,124) .. controls (160.5,121.51) and (162.51,119.5) .. (165,119.5) .. controls (167.49,119.5) and (169.5,121.51) .. (169.5,124) .. controls (169.5,126.49) and (167.49,128.5) .. (165,128.5) .. controls (162.51,128.5) and (160.5,126.49) .. (160.5,124) -- cycle ;
\draw   (334,169.5) .. controls (334,114.55) and (378.55,70) .. (433.5,70) .. controls (488.45,70) and (533,114.55) .. (533,169.5) .. controls (533,224.45) and (488.45,269) .. (433.5,269) .. controls (378.55,269) and (334,224.45) .. (334,169.5) -- cycle ;
\draw    (433.5,269) .. controls (416,267) and (413,81) .. (433.5,70) ;
\draw  [dash pattern={on 4.5pt off 4.5pt}]  (433.5,70) .. controls (455,73) and (453,269) .. (433.5,269) ;
\draw [color={rgb, 255:red, 144; green, 19; blue, 254 }  ,draw opacity=1 ][line width=3]    (421,122) .. controls (419,134) and (416,198) .. (421,215) ;
\draw  [fill={rgb, 255:red, 0; green, 0; blue, 0 }  ,fill opacity=1 ] (416.5,126.5) .. controls (416.5,124.01) and (418.51,122) .. (421,122) .. controls (423.49,122) and (425.5,124.01) .. (425.5,126.5) .. controls (425.5,128.99) and (423.49,131) .. (421,131) .. controls (418.51,131) and (416.5,128.99) .. (416.5,126.5) -- cycle ;
\draw  [fill={rgb, 255:red, 0; green, 0; blue, 0 }  ,fill opacity=1 ] (416.5,215) .. controls (416.5,212.51) and (418.51,210.5) .. (421,210.5) .. controls (423.49,210.5) and (425.5,212.51) .. (425.5,215) .. controls (425.5,217.49) and (423.49,219.5) .. (421,219.5) .. controls (418.51,219.5) and (416.5,217.49) .. (416.5,215) -- cycle ;

\draw (42,151.4) node [anchor=north west][inner sep=0.75pt]    {$\bra{\varphi }$};
\draw (227,159.4) node [anchor=north west][inner sep=0.75pt]    {$\ket{\psi }$};
\draw (103,153.4) node [anchor=north west][inner sep=0.75pt]  [color={rgb, 255:red, 144; green, 19; blue, 254 }  ,opacity=1 ]  {$A$};
\draw (169,154.4) node [anchor=north west][inner sep=0.75pt]  [color={rgb, 255:red, 144; green, 19; blue, 254 }  ,opacity=1 ]  {$A$};
\draw (396,161.4) node [anchor=north west][inner sep=0.75pt]  [color={rgb, 255:red, 144; green, 19; blue, 254 }  ,opacity=1 ]  {$A$};

\end{tikzpicture}
\end{center}

\caption{The subsystem $A$ within a non-unitary QFT residing on a two-sphere, with black points indicating the codimension-2 entangling surface.}\label{img2}
\end{figure}

First, we briefly introduce the pseudoentropy. Dividing the total
system into two subsystems $A$ and $B$, the pseudoentropy is defined
by the von Neumann entropy, 
\begin{equation}
S_{A}=-\mathrm{Tr}\left[\tau_{A}\log\tau_{A}\right],
\end{equation}
of the reduced transition matrix 
\begin{equation}
\tau_{A}=\mathrm{Tr}_{B}\left[\frac{\left|\psi\right\rangle \left\langle \varphi\right|}{\left\langle \varphi\mid\psi\right\rangle }\right],
\end{equation}
where $\left|\psi\right\rangle $ and $\left|\varphi\right\rangle $
are two different quantum states in the total Hilbert space that is
factorized as $\mathcal{H}=\mathcal{H}_{A}\otimes\mathcal{H}_{B}$.
For a generic QFT living on a curved surface $\Sigma$, the pseudoentropy
could be captured by the replica method \cite{Calabrese:2004eu,Calabrese:2009qy}
in path integral formalism. Denoting the manifold corresponding to
$\left\langle \varphi\mid\psi\right\rangle $ as $\Sigma$ and the
manifold corresponding to $\mathrm{Tr}_{A}\left(\tau_{A}\right)^{n}$
as $\Sigma_{n}$, the pseudoentropy for the subsystem $A$ reads
\begin{equation}
S_{A}=\lim_{n\rightarrow1}\frac{1}{1-n}\log\left[\frac{Z_{\Sigma_{n}}}{\left(Z_{\Sigma}\right)^{n}}\right],\label{eq:entropy}
\end{equation}
where $Z_{\Sigma}$ is the path integral over the manifold $\Sigma$
and $S_{A}$ can be regarded as a well-defined entanglement entropy
in the dS$_{3}$ context. As an example, one can compute the pseudoentropy for subsystem $A$, which corresponds to an interval, in a non-unitary QFT residing on a sphere $\mathbb{S}^{2}$, as depicted in Figure \ref{img2}.

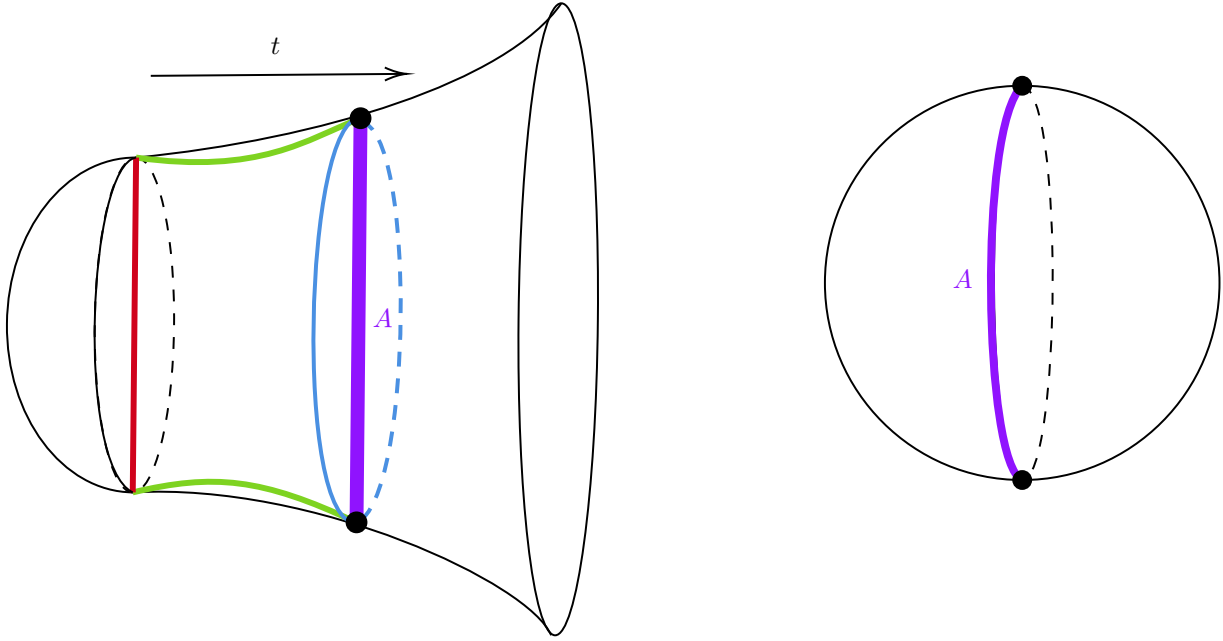
\begin{figure}

\tikzset{every picture/.style={line width=0.75pt}} 

\begin{tikzpicture}[x=0.75pt,y=0.75pt,yscale=-1,xscale=1]

\draw  [draw opacity=0] (132.91,308.25) .. controls (132.91,308.25) and (132.91,308.25) .. (132.91,308.25) .. controls (97.4,307.9) and (68.99,269.78) .. (69.46,223.12) .. controls (69.92,176.45) and (99.09,138.91) .. (134.6,139.26) -- (133.76,223.76) -- cycle ; \draw   (132.91,308.25) .. controls (132.91,308.25) and (132.91,308.25) .. (132.91,308.25) .. controls (97.4,307.9) and (68.99,269.78) .. (69.46,223.12) .. controls (69.92,176.45) and (99.09,138.91) .. (134.6,139.26) ;  
\draw    (349.08,61.39) .. controls (321.68,101.12) and (227.39,130.19) .. (134.6,139.26) ;
\draw    (343.91,380.36) .. controls (333.17,354.25) and (224.67,303.17) .. (132.91,308.25) ;
\draw  [dash pattern={on 4.5pt off 4.5pt}] (134.6,139.26) .. controls (145.64,139.37) and (154.22,177.29) .. (153.75,223.96) .. controls (153.29,270.62) and (143.96,308.36) .. (132.91,308.25) .. controls (121.87,308.14) and (113.29,270.23) .. (113.76,223.56) .. controls (114.22,176.89) and (123.55,139.15) .. (134.6,139.26) -- cycle ;
\draw [color={rgb, 255:red, 144; green, 19; blue, 254 }  ,draw opacity=1 ][line width=5.25]    (247.8,119.39) -- (245.77,323.38) ;
\draw [color={rgb, 255:red, 126; green, 211; blue, 33 }  ,draw opacity=1 ][line width=2.25]    (134.6,139.26) .. controls (197.31,147.89) and (220.49,130.12) .. (247.8,119.39) ;
\draw [color={rgb, 255:red, 126; green, 211; blue, 33 }  ,draw opacity=1 ][line width=2.25]    (132.91,308.25) .. controls (173.81,298.66) and (197.79,300.9) .. (245.77,323.38) ;
\draw [color={rgb, 255:red, 208; green, 2; blue, 27 }  ,draw opacity=1 ][line width=2.25]    (134.6,139.26) -- (132.91,308.25) ;
\draw [color={rgb, 255:red, 74; green, 144; blue, 226 }  ,draw opacity=1 ][line width=1.5]    (247.8,119.39) .. controls (219.11,118.1) and (214.04,326.06) .. (245.77,323.38) ;
\draw    (134.6,139.26) .. controls (110.36,143.02) and (103.86,293.96) .. (132.91,308.25) ;
\draw [color={rgb, 255:red, 74; green, 144; blue, 226 }  ,draw opacity=1 ][line width=1.5]  [dash pattern={on 5.63pt off 4.5pt}]  (247.78,121.39) .. controls (276.06,123.67) and (274.18,312.66) .. (245.77,323.38) ;
\draw   (349.08,61.39) .. controls (360.13,61.5) and (368.37,133) .. (367.49,221.08) .. controls (366.62,309.17) and (356.95,380.49) .. (345.91,380.38) .. controls (334.86,380.27) and (326.62,308.77) .. (327.49,220.69) .. controls (328.37,132.6) and (338.04,61.28) .. (349.08,61.39) -- cycle ;
\draw  [fill={rgb, 255:red, 0; green, 0; blue, 0 }  ,fill opacity=1 ] (247.85,114.39) .. controls (250.61,114.42) and (252.83,116.68) .. (252.8,119.44) .. controls (252.77,122.2) and (250.51,124.42) .. (247.75,124.39) .. controls (244.99,124.36) and (242.77,122.1) .. (242.8,119.34) .. controls (242.83,116.58) and (245.09,114.36) .. (247.85,114.39) -- cycle ;
\draw  [fill={rgb, 255:red, 0; green, 0; blue, 0 }  ,fill opacity=1 ] (245.82,318.38) .. controls (248.58,318.41) and (250.8,320.67) .. (250.77,323.43) .. controls (250.74,326.19) and (248.48,328.41) .. (245.72,328.38) .. controls (242.96,328.35) and (240.74,326.09) .. (240.77,323.33) .. controls (240.8,320.57) and (243.06,318.35) .. (245.82,318.38) -- cycle ;
\draw    (142,98) -- (269,97.02) ;
\draw [shift={(271,97)}, rotate = 179.56] [color={rgb, 255:red, 0; green, 0; blue, 0 }  ][line width=0.75]    (10.93,-3.29) .. controls (6.95,-1.4) and (3.31,-0.3) .. (0,0) .. controls (3.31,0.3) and (6.95,1.4) .. (10.93,3.29)   ;
\draw   (482,202.5) .. controls (482,147.55) and (526.55,103) .. (581.5,103) .. controls (636.45,103) and (681,147.55) .. (681,202.5) .. controls (681,257.45) and (636.45,302) .. (581.5,302) .. controls (526.55,302) and (482,257.45) .. (482,202.5) -- cycle ;
\draw    (581.5,302) .. controls (564,300) and (561,114) .. (581.5,103) ;
\draw  [dash pattern={on 4.5pt off 4.5pt}]  (581.5,103) .. controls (603,106) and (601,302) .. (581.5,302) ;
\draw [color={rgb, 255:red, 144; green, 19; blue, 254 }  ,draw opacity=1 ][line width=3]    (581.5,103) .. controls (559,124) and (562,294) .. (581.5,302) ;
\draw  [fill={rgb, 255:red, 0; green, 0; blue, 0 }  ,fill opacity=1 ] (577,103) .. controls (577,100.51) and (579.01,98.5) .. (581.5,98.5) .. controls (583.99,98.5) and (586,100.51) .. (586,103) .. controls (586,105.49) and (583.99,107.5) .. (581.5,107.5) .. controls (579.01,107.5) and (577,105.49) .. (577,103) -- cycle ;
\draw  [fill={rgb, 255:red, 0; green, 0; blue, 0 }  ,fill opacity=1 ] (577,302) .. controls (577,299.51) and (579.01,297.5) .. (581.5,297.5) .. controls (583.99,297.5) and (586,299.51) .. (586,302) .. controls (586,304.49) and (583.99,306.5) .. (581.5,306.5) .. controls (579.01,306.5) and (577,304.49) .. (577,302) -- cycle ;

\draw (252.55,214.32) node [anchor=north west][inner sep=0.75pt]  [color={rgb, 255:red, 144; green, 19; blue, 254 }  ,opacity=1 ,rotate=-0.65]  {$A$};
\draw (201,77.4) node [anchor=north west][inner sep=0.75pt]    {$t$};
\draw (545,194.4) node [anchor=north west][inner sep=0.75pt]  [color={rgb, 255:red, 144; green, 19; blue, 254 }  ,opacity=1 ]  {$A$};

\end{tikzpicture}
\caption{Left panel: Geodesics connecting to the entangling surface in the $\text{dS}_{3}$.
The red line denotes one spacelike geodesic and two green lines denote
two timelike geodesics. The purple line denotes a spacelike interval
$A$ on the two-sphere. Right panel: The entangling surface consists of two antipodal points on the two-sphere.}
\label{fig3}
\end{figure}

To capture the pseudoentropy, we first need to calculate the partition
function for a field theory. Specifically, one saddle solution for the
Hartle-Hawking wavefunction $\Psi_{\text{dS}}\left[\gamma\right]$
is the Euclidean sphere $\mathbb{S}^{2}$, where the corresponding
metric of de Sitter spacetime is given by
\begin{equation}
ds^{2}=\ell_{\text{dS}}^{2}\left(-dt^{2}+\cosh^{2}t\,d\Omega_{2}^{2}\right),
\end{equation}
where $d\Omega_{2}^{2}=d\theta^{2}+\sin^{2}\theta\,d\phi^{2}$ is
the metric of a $2d$ unit sphere, and the spacelike boundary at  
time $t$ is a Euclidean sphere $\mathbb{S}^{2}$ with the radius
$r=\sqrt{\frac{\lambda_{\text{dS}}\,c_{\text{dS}}}{12}}\cosh t$.
In this section, we will calculate the pseudoentropy of the $T\bar{T}$
deformed field theory living on a sphere $\mathbb{S}^{2}$ with a
radius $r$. 
To be precise, we focus on the case that an entangling
surface consists of two antipodal points on this sphere, as shown in the right panel in Figure \ref{fig3}.  Following
\cite{Donnelly:2018bef}, for a field theory living on a sphere
with the metric
\begin{equation}
ds^{2}=r^{2}\left(d\theta^{2}+\sin^{2}\theta\,d\phi^{2}\right),
\end{equation}
the stress tensor takes the form\footnote{Generally, for vacuum states in QFTs living in a $d$-dimensional  maximally symmetric space, the stress tensor satisfies $\left\langle T_{\mu\nu}(x)\right\rangle =\frac{1}{d}\left\langle \Theta\right\rangle g_{\mu\nu}(x)$, where $\left\langle \Theta\right\rangle$ is an $x$-independent constant.}
\begin{equation}
T_{ab}=\alpha\gamma_{ab},\label{eq:tensor}
\end{equation}
where $\alpha$ could be determined by substituting eqn.(\ref{eq:tensor}) into the trace flow equation (\ref{eq:traceflow}),
\begin{equation}
\alpha=\frac{1}{\pi\lambda}\left(1-\sqrt{1+\frac{\lambda c}{12r^{2}}}\right)=\frac{\text{i}}{\pi\lambda_{\text{dS}}}\left(1-\sqrt{1-\frac{\lambda_{\text{dS}}\,c_{\text{dS}}}{12r^{2}}}\right).
\end{equation}
Noticing that $r\partial_{r}\gamma_{ab}=2\gamma_{ab}$, one obtains
the equation for the partition function 
\begin{equation}
\frac{d\log Z_{\text{QFT}}}{dr}=-\frac{1}{r}\int_{\Sigma}d^{2}x\sqrt{\gamma}\,T_{a}^{a}=\text{i}\frac{8}{\lambda_{\text{dS}}}\left(\sqrt{r^{2}-\frac{\lambda_{\text{dS}}\,c_{\text{dS}}}{12}}-r\right),
\end{equation}
and the partition function thus reads
\begin{equation}
\log Z_{\text{QFT}}=\text{i}\frac{4}{\lambda_{\text{dS}}}\left[r\left(\sqrt{r^{2}-\frac{\lambda_{\text{dS}}\,c_{\text{dS}}}{12}}-r\right)-\frac{\lambda_{\text{dS}}\,c_{\text{dS}}}{12}\tanh^{-1}\left(\frac{r}{\sqrt{r^{2}-\lambda_{\text{dS}}\,c_{\text{dS}}/12}}\right)\right].
\end{equation}
It is worthy to note that, since $r/\sqrt{r^{2}-\lambda_{\text{dS}}\,c_{\text{dS}}/12}>1$,
the function $\tanh^{-1}\left(r/\sqrt{r^{2}-\lambda_{\text{dS}}\,c_{\text{dS}}/12}\right)$
is indeed complex-valued. Focusing on the principal branch of the
inverse hyperbolic function, one then obtains 
\begin{equation}
\log Z_{\text{QFT}}=\text{i}\frac{4}{\lambda_{\text{dS}}}\left[r\left(\sqrt{r^{2}-\frac{\lambda_{\text{dS}}\,c_{\text{dS}}}{12}}-r\right)-\frac{\lambda_{\text{dS}}\,c_{\text{dS}}}{12}\coth^{-1}\left(\frac{r}{\sqrt{r^{2}-\lambda_{\text{dS}}\,c_{\text{dS}}/12}}\right)\right]+\frac{\pi c_{\text{dS}}}{6}.
\end{equation}
The real part $\frac{\pi c_{\text{dS}}}{6}$ is consistent with the
result in \cite{Hikida:2022ltr}:
\begin{equation}
\left|Z_{\text{CFT}}\right|^{2}=\exp\left(\frac{\pi c_{\text{dS}}}{3}\right)
\label{eq: Partition CFT}
\end{equation}
since the deformation parameter is imaginary-valued and the $T\bar{T}$
deformation only affects the imaginary part of $\log Z$.

Utilizing the replica method introduced in \cite{Donnelly:2018bef}, in the case where the entangling surface consists of two antipodal points on the two-sphere, the $n$-sheeted cover is simply
\begin{equation}
ds^{2}=g_{ab}dx^a dx^b =r^{2}\left(d\theta^{2}+n^2\sin^{2}\theta\,d\phi^{2}\right),
\end{equation}
and the pseudoentropy reads
\begin{equation}
S_{A}=\lim_{n\rightarrow1}\frac{1}{1-n}\log\left[\frac{Z_{n}}{\left(Z_{1}\right)^{n}}\right]=\left.\left(1-n\frac{\partial}{\partial n}\right)\log Z_{n}\right|_{n=1}.
\end{equation}
Noting that $n\partial_{n}g_{\phi\phi}=2g_{\phi\phi}$, the variation of $\log Z_n$ with respect to $n$ can be expressed as
\begin{equation}
\left. n\frac{\partial}{\partial n}\log Z_{n}\right|_{n=1}=-\frac{1}{2}\int_{\Sigma}d^{2}x\sqrt{\gamma}\,T_{a}^{a}=\frac{r}{2}\frac{d}{dr}\log Z_{\text{QFT}}.
\end{equation}

We could thus obtain the pseudoentropy $S_{A}$ for an entangling surface
of two antipodal points on a sphere $\mathbb{S}^{2}$:
\begin{equation}
S_A=\left(1-\frac{r}{2}\frac{d}{dr}\right)\log Z_{\text{QFT}}=-\text{i}\,\frac{c_{\text{dS}}}{3}\coth^{-1}\left(\frac{r}{\sqrt{r^{2}-\lambda_{\text{dS}}\,c_{\text{dS}}/12}}\right)+\frac{\pi c_{\text{dS}}}{6}.
\end{equation}
Substituting $r=\sqrt{\frac{\lambda_{\text{dS}}\,c_{\text{dS}}}{12}}\cosh t$,
the entanglement entropy of two antipodal points on a $\mathbb{S}^{2}$
is thus given by 
\begin{equation}
S_{A}=\frac{c_{\text{dS}}}{6}\pi-\text{i}\,\frac{c_{\text{dS}}}{3}t.
\label{eq: AntiEE}
\end{equation}
On the other hand, in the $\text{dS}_{3}$ bulk, the geodesic distance between two points
at the same time $\left(t,\theta_{i}=0,\phi=0\right)$ and $\left(t,\theta_{f}=\pi,\phi=0\right)$
is given by
\begin{equation}
D\left(\theta_{i},\theta_{f}\right)=\ell_{\text{dS}}\mathrm{arcos}\left[1-2\cosh^{2}t\sin^{2}\left(\frac{\theta_{f}-\theta_{i}}{2}\right)\right]=\ell_{\text{dS}}\pi-2\,\text{i}\,\ell_{\text{dS}}t.
\end{equation}
According to the Ryu-Takayanagi formula \cite{Ryu:2006bv}, Refs. \cite{Hikida:2022ltr, Doi:2022iyj} have determined that the complex-valued extremal surface comprises one spacelike geodesic and two timelike geodesics, as illustrated in Figure \ref{fig3}. The two timelike geodesics connect the entangling surface and the de Sitter horizon, respectively, while the spacelike geodesic links the endpoints of the two timelike geodesics on the de Sitter horizon. Furthermore, the length of the spacelike geodesic is proportional to the real part of the pseudoentropy, whereas the total length of the two timelike geodesics is proportional to the imaginary part of the pseudoentropy. Related works on the complex-valued extremal surface have also been proposed in \cite{Narayan:2015vda, Narayan:2015oka, Narayan:2016xwq}. Using the identifications eqn.(\ref{eq:relation}), the RT
formula gives
\begin{equation}
S_{A}=\frac{D\left(\theta_{i},\theta_{f}\right)}{4G_{N}}=\frac{c_{\text{dS}}}{6}\pi-\text{i}\,\frac{c_{\text{dS}}}{3}t,
\end{equation}
which exactly is equal to the  entanglement entropy eqn.(\ref{eq: AntiEE}). 

Therefore, as promised, we verified that, for a static finite volume slice  $\mathbb{S}^{2}$ in dS$_3$, 
the pseudoentropy for an entangling surface consisting of two antipodal points is
precisely equal to the complex geodesic in the $\text{dS}_{3}$ bulk.


%

\vspace{3ex} 
\noindent {\bf Acknowledgements} 
This work is supported in part by NSFC (Grant No. 12275184 and 11875196). 

\appendix

\bibliographystyle{unsrturl}
\bibliography{new}

\begin{thebibliography}{10}

\bibitem{Strominger:2001pn}
Andrew Strominger.
\newblock {The dS / CFT correspondence}.
\newblock {\em JHEP}, 10:034, 2001.
\newblock \href {http://arxiv.org/abs/hep-th/0106113}
  {\path{arXiv:hep-th/0106113}}, \href
  {https://doi.org/10.1088/1126-6708/2001/10/034}
  {\path{doi:10.1088/1126-6708/2001/10/034}}.

\bibitem{Maldacena:1997re}
Juan~Martin Maldacena.
\newblock {The Large N limit of superconformal field theories and
  supergravity}.
\newblock {\em Adv. Theor. Math. Phys.}, 2:231--252, 1998.
\newblock \href {http://arxiv.org/abs/hep-th/9711200}
  {\path{arXiv:hep-th/9711200}}, \href
  {https://doi.org/10.1023/A:1026654312961}
  {\path{doi:10.1023/A:1026654312961}}.

\bibitem{Witten:1998qj}
Edward Witten.
\newblock {Anti-de Sitter space and holography}.
\newblock {\em Adv. Theor. Math. Phys.}, 2:253--291, 1998.
\newblock \href {http://arxiv.org/abs/hep-th/9802150}
  {\path{arXiv:hep-th/9802150}}, \href
  {https://doi.org/10.4310/ATMP.1998.v2.n2.a2}
  {\path{doi:10.4310/ATMP.1998.v2.n2.a2}}.

\bibitem{Hikida:2021ese}
Yasuaki Hikida, Tatsuma Nishioka, Tadashi Takayanagi, and Yusuke Taki.
\newblock {Holography in de Sitter Space via Chern-Simons Gauge Theory}.
\newblock {\em Phys. Rev. Lett.}, 129(4):041601, 2022.
\newblock \href {http://arxiv.org/abs/2110.03197} {\path{arXiv:2110.03197}},
  \href {https://doi.org/10.1103/PhysRevLett.129.041601}
  {\path{doi:10.1103/PhysRevLett.129.041601}}.

\bibitem{Hikida:2022ltr}
Yasuaki Hikida, Tatsuma Nishioka, Tadashi Takayanagi, and Yusuke Taki.
\newblock {CFT duals of three-dimensional de Sitter gravity}.
\newblock {\em JHEP}, 05:129, 2022.
\newblock \href {http://arxiv.org/abs/2203.02852} {\path{arXiv:2203.02852}},
  \href {https://doi.org/10.1007/JHEP05(2022)129}
  {\path{doi:10.1007/JHEP05(2022)129}}.

\bibitem{Zamolodchikov:2004ce}
Alexander~B. Zamolodchikov.
\newblock {Expectation value of composite field T anti-T in two-dimensional
  quantum field theory}.
\newblock 1 2004.
\newblock \href {http://arxiv.org/abs/hep-th/0401146}
  {\path{arXiv:hep-th/0401146}}.

\bibitem{Cavaglia:2016oda}
Andrea Cavagli\`a, Stefano Negro, Istv\'an~M. Sz\'ecs\'enyi, and Roberto Tateo.
\newblock {$T \bar{T}$-deformed 2D Quantum Field Theories}.
\newblock {\em JHEP}, 10:112, 2016.
\newblock \href {http://arxiv.org/abs/1608.05534} {\path{arXiv:1608.05534}},
  \href {https://doi.org/10.1007/JHEP10(2016)112}
  {\path{doi:10.1007/JHEP10(2016)112}}.

\bibitem{Smirnov:2016lqw}
F.~A. Smirnov and A.~B. Zamolodchikov.
\newblock {On space of integrable quantum field theories}.
\newblock {\em Nucl. Phys. B}, 915:363--383, 2017.
\newblock \href {http://arxiv.org/abs/1608.05499} {\path{arXiv:1608.05499}},
  \href {https://doi.org/10.1016/j.nuclphysb.2016.12.014}
  {\path{doi:10.1016/j.nuclphysb.2016.12.014}}.

\bibitem{Dubovsky:2017cnj}
Sergei Dubovsky, Victor Gorbenko, and Mehrdad Mirbabayi.
\newblock {Asymptotic fragility, near AdS$_{2}$ holography and $ T\overline{T}
  $}.
\newblock {\em JHEP}, 09:136, 2017.
\newblock \href {http://arxiv.org/abs/1706.06604} {\path{arXiv:1706.06604}},
  \href {https://doi.org/10.1007/JHEP09(2017)136}
  {\path{doi:10.1007/JHEP09(2017)136}}.

\bibitem{Donnelly:2018bef}
William Donnelly and Vasudev Shyam.
\newblock {Entanglement entropy and $T \overline{T}$ deformation}.
\newblock {\em Phys. Rev. Lett.}, 121(13):131602, 2018.
\newblock \href {http://arxiv.org/abs/1806.07444} {\path{arXiv:1806.07444}},
  \href {https://doi.org/10.1103/PhysRevLett.121.131602}
  {\path{doi:10.1103/PhysRevLett.121.131602}}.

\bibitem{Kraus:2018xrn}
Per Kraus, Junyu Liu, and Donald Marolf.
\newblock {Cutoff AdS$_{3}$ versus the $ T\overline{T} $ deformation}.
\newblock {\em JHEP}, 07:027, 2018.
\newblock \href {http://arxiv.org/abs/1801.02714} {\path{arXiv:1801.02714}},
  \href {https://doi.org/10.1007/JHEP07(2018)027}
  {\path{doi:10.1007/JHEP07(2018)027}}.

\bibitem{Aharony:2018vux}
Ofer Aharony and Talya Vaknin.
\newblock {The TT* deformation at large central charge}.
\newblock {\em JHEP}, 05:166, 2018.
\newblock \href {http://arxiv.org/abs/1803.00100} {\path{arXiv:1803.00100}},
  \href {https://doi.org/10.1007/JHEP05(2018)166}
  {\path{doi:10.1007/JHEP05(2018)166}}.

\bibitem{Conti:2018tca}
Riccardo Conti, Stefano Negro, and Roberto Tateo.
\newblock {The $ \mathrm{T}\overline{\mathrm{T}} $ perturbation and its
  geometric interpretation}.
\newblock {\em JHEP}, 02:085, 2019.
\newblock \href {http://arxiv.org/abs/1809.09593} {\path{arXiv:1809.09593}},
  \href {https://doi.org/10.1007/JHEP02(2019)085}
  {\path{doi:10.1007/JHEP02(2019)085}}.

\bibitem{Cardy:2018sdv}
John Cardy.
\newblock {The $ T\overline{T} $ deformation of quantum field theory as random
  geometry}.
\newblock {\em JHEP}, 10:186, 2018.
\newblock \href {http://arxiv.org/abs/1801.06895} {\path{arXiv:1801.06895}},
  \href {https://doi.org/10.1007/JHEP10(2018)186}
  {\path{doi:10.1007/JHEP10(2018)186}}.

\bibitem{Conti:2018jho}
Riccardo Conti, Leonardo Iannella, Stefano Negro, and Roberto Tateo.
\newblock {Generalised Born-Infeld models, Lax operators and the $
  \mathrm{T}\overline{\mathrm{T}} $ perturbation}.
\newblock {\em JHEP}, 11:007, 2018.
\newblock \href {http://arxiv.org/abs/1806.11515} {\path{arXiv:1806.11515}},
  \href {https://doi.org/10.1007/JHEP11(2018)007}
  {\path{doi:10.1007/JHEP11(2018)007}}.

\bibitem{Bonelli:2018kik}
Giulio Bonelli, Nima Doroud, and Mengqi Zhu.
\newblock {$T \bar{T}$-deformations in closed form}.
\newblock {\em JHEP}, 06:149, 2018.
\newblock \href {http://arxiv.org/abs/1804.10967} {\path{arXiv:1804.10967}},
  \href {https://doi.org/10.1007/JHEP06(2018)149}
  {\path{doi:10.1007/JHEP06(2018)149}}.

\bibitem{Aharony:2018bad}
Ofer Aharony, Shouvik Datta, Amit Giveon, Yunfeng Jiang, and David Kutasov.
\newblock {Modular invariance and uniqueness of $T\bar{T}$ deformed CFT}.
\newblock {\em JHEP}, 01:086, 2019.
\newblock \href {http://arxiv.org/abs/1808.02492} {\path{arXiv:1808.02492}},
  \href {https://doi.org/10.1007/JHEP01(2019)086}
  {\path{doi:10.1007/JHEP01(2019)086}}.

\bibitem{Datta:2018thy}
Shouvik Datta and Yunfeng Jiang.
\newblock {$T\bar{T}$ deformed partition functions}.
\newblock {\em JHEP}, 08:106, 2018.
\newblock \href {http://arxiv.org/abs/1806.07426} {\path{arXiv:1806.07426}},
  \href {https://doi.org/10.1007/JHEP08(2018)106}
  {\path{doi:10.1007/JHEP08(2018)106}}.

\bibitem{Chen:2019mis}
Bin Chen, Lin Chen, and Cheng-Yong Zhang.
\newblock {Surface/state correspondence and $T\overline{T}$ deformation}.
\newblock {\em Phys. Rev. D}, 101(10):106011, 2020.
\newblock \href {http://arxiv.org/abs/1907.12110} {\path{arXiv:1907.12110}},
  \href {https://doi.org/10.1103/PhysRevD.101.106011}
  {\path{doi:10.1103/PhysRevD.101.106011}}.

\bibitem{Conti:2019dxg}
Riccardo Conti, Stefano Negro, and Roberto Tateo.
\newblock {Conserved currents and $\text{T}\bar{\text{T}}_s$ irrelevant
  deformations of 2D integrable field theories}.
\newblock {\em JHEP}, 11:120, 2019.
\newblock \href {http://arxiv.org/abs/1904.09141} {\path{arXiv:1904.09141}},
  \href {https://doi.org/10.1007/JHEP11(2019)120}
  {\path{doi:10.1007/JHEP11(2019)120}}.

\bibitem{Guica:2019nzm}
Monica Guica and Ruben Monten.
\newblock {$T\bar T$ and the mirage of a bulk cutoff}.
\newblock {\em SciPost Phys.}, 10(2):024, 2021.
\newblock \href {http://arxiv.org/abs/1906.11251} {\path{arXiv:1906.11251}},
  \href {https://doi.org/10.21468/SciPostPhys.10.2.024}
  {\path{doi:10.21468/SciPostPhys.10.2.024}}.

\bibitem{Ishii:2019uwk}
Takaaki Ishii, Suguru Okumura, Jun-Ichi Sakamoto, and Kentaroh Yoshida.
\newblock {Gravitational perturbations as $T\bar{T}$-deformations in 2D dilaton
  gravity systems}.
\newblock {\em Nucl. Phys. B}, 951:114901, 2020.
\newblock \href {http://arxiv.org/abs/1906.03865} {\path{arXiv:1906.03865}},
  \href {https://doi.org/10.1016/j.nuclphysb.2019.114901}
  {\path{doi:10.1016/j.nuclphysb.2019.114901}}.

\bibitem{Grieninger:2019zts}
Sebastian Grieninger.
\newblock {Entanglement entropy and $ T\overline{T} $ deformations beyond
  antipodal points from holography}.
\newblock {\em JHEP}, 11:171, 2019.
\newblock \href {http://arxiv.org/abs/1908.10372} {\path{arXiv:1908.10372}},
  \href {https://doi.org/10.1007/JHEP11(2019)171}
  {\path{doi:10.1007/JHEP11(2019)171}}.

\bibitem{Jeong:2019ylz}
Hyun-Sik Jeong, Keun-Young Kim, and Mitsuhiro Nishida.
\newblock {Entanglement and R\'enyi entropy of multiple intervals in
  $T\overline{T}$-deformed CFT and holography}.
\newblock {\em Phys. Rev. D}, 100(10):106015, 2019.
\newblock \href {http://arxiv.org/abs/1906.03894} {\path{arXiv:1906.03894}},
  \href {https://doi.org/10.1103/PhysRevD.100.106015}
  {\path{doi:10.1103/PhysRevD.100.106015}}.

\bibitem{Jiang:2019epa}
Yunfeng Jiang.
\newblock {A pedagogical review on solvable irrelevant deformations of 2D
  quantum field theory}.
\newblock {\em Commun. Theor. Phys.}, 73(5):057201, 2021.
\newblock \href {http://arxiv.org/abs/1904.13376} {\path{arXiv:1904.13376}},
  \href {https://doi.org/10.1088/1572-9494/abe4c9}
  {\path{doi:10.1088/1572-9494/abe4c9}}.

\bibitem{He:2019vzf}
Song He and Hongfei Shu.
\newblock {Correlation functions, entanglement and chaos in the $
  T\overline{T}/J\overline{T} $-deformed CFTs}.
\newblock {\em JHEP}, 02:088, 2020.
\newblock \href {http://arxiv.org/abs/1907.12603} {\path{arXiv:1907.12603}},
  \href {https://doi.org/10.1007/JHEP02(2020)088}
  {\path{doi:10.1007/JHEP02(2020)088}}.

\bibitem{Jiang:2019tcq}
Yunfeng Jiang.
\newblock {Expectation value of $\mathrm{T}\overline{\mathrm{T}}$ operator in
  curved spacetimes}.
\newblock {\em JHEP}, 02:094, 2020.
\newblock \href {http://arxiv.org/abs/1903.07561} {\path{arXiv:1903.07561}},
  \href {https://doi.org/10.1007/JHEP02(2020)094}
  {\path{doi:10.1007/JHEP02(2020)094}}.

\bibitem{Pozsgay:2019ekd}
Bal\'azs Pozsgay, Yunfeng Jiang, and G\'abor Tak\'acs.
\newblock {$T\bar T$-deformation and long range spin chains}.
\newblock {\em JHEP}, 03:092, 2020.
\newblock \href {http://arxiv.org/abs/1911.11118} {\path{arXiv:1911.11118}},
  \href {https://doi.org/10.1007/JHEP03(2020)092}
  {\path{doi:10.1007/JHEP03(2020)092}}.

\bibitem{Grado-White:2020wlb}
Brianna Grado-White, Donald Marolf, and Sean~J. Weinberg.
\newblock {Radial Cutoffs and Holographic Entanglement}.
\newblock {\em JHEP}, 01:009, 2021.
\newblock \href {http://arxiv.org/abs/2008.07022} {\path{arXiv:2008.07022}},
  \href {https://doi.org/10.1007/JHEP01(2021)009}
  {\path{doi:10.1007/JHEP01(2021)009}}.

\bibitem{Khoeini-Moghaddam:2020ymm}
Salomeh Khoeini-Moghaddam, Farzad Omidi, and Chandrima Paul.
\newblock {Aspects of Hyperscaling Violating Geometries at Finite Cutoff}.
\newblock {\em JHEP}, 02:121, 2021.
\newblock \href {http://arxiv.org/abs/2011.00305} {\path{arXiv:2011.00305}},
  \href {https://doi.org/10.1007/JHEP02(2021)121}
  {\path{doi:10.1007/JHEP02(2021)121}}.

\bibitem{Caputa:2020lpa}
Pawel Caputa, Pawel Caputa, Shouvik Datta, Shouvik Datta, Yunfeng Jiang,
  Yunfeng Jiang, Per Kraus, and Per Kraus.
\newblock {Geometrizing $ T\overline{T} $}.
\newblock {\em JHEP}, 03:140, 2021.
\newblock [Erratum: JHEP 09, 110 (2022)].
\newblock \href {http://arxiv.org/abs/2011.04664} {\path{arXiv:2011.04664}},
  \href {https://doi.org/10.1007/JHEP03(2021)140}
  {\path{doi:10.1007/JHEP03(2021)140}}.

\bibitem{Medenjak:2020ppv}
Marko Medenjak, Giuseppe Policastro, and Takato Yoshimura.
\newblock {$T\bar{T}$-Deformed Conformal Field Theories out of Equilibrium}.
\newblock {\em Phys. Rev. Lett.}, 126(12):121601, 2021.
\newblock \href {http://arxiv.org/abs/2011.05827} {\path{arXiv:2011.05827}},
  \href {https://doi.org/10.1103/PhysRevLett.126.121601}
  {\path{doi:10.1103/PhysRevLett.126.121601}}.

\bibitem{Li:2020pwa}
Yi~Li and Yang Zhou.
\newblock {Cutoff AdS$_{3}$ versus $ T\overline{T} $ CFT$_{2}$ in the large
  central charge sector: correlators of energy-momentum tensor}.
\newblock {\em JHEP}, 12:168, 2020.
\newblock \href {http://arxiv.org/abs/2005.01693} {\path{arXiv:2005.01693}},
  \href {https://doi.org/10.1007/JHEP12(2020)168}
  {\path{doi:10.1007/JHEP12(2020)168}}.

\bibitem{He:2020qcs}
Song He.
\newblock {Note on higher-point correlation functions of the $T\bar T$ or
  $J\bar T$ deformed CFTs}.
\newblock {\em Sci. China Phys. Mech. Astron.}, 64(9):291011, 2021.
\newblock \href {http://arxiv.org/abs/2012.06202} {\path{arXiv:2012.06202}},
  \href {https://doi.org/10.1007/s11433-021-1741-1}
  {\path{doi:10.1007/s11433-021-1741-1}}.

\bibitem{Ceschin:2020jto}
Paolo Ceschin, Riccardo Conti, and Roberto Tateo.
\newblock {$ \mathrm{T}\overline{\mathrm{T}} $-deformed nonlinear
  Schr\"odinger}.
\newblock {\em JHEP}, 04:121, 2021.
\newblock \href {http://arxiv.org/abs/2012.12760} {\path{arXiv:2012.12760}},
  \href {https://doi.org/10.1007/JHEP04(2021)121}
  {\path{doi:10.1007/JHEP04(2021)121}}.

\bibitem{Jiang:2020nnb}
Yunfeng Jiang.
\newblock {$\mathrm{T}\overline{\mathrm{T}}$-deformed 1d Bose gas}.
\newblock {\em SciPost Phys.}, 12(6):191, 2022.
\newblock \href {http://arxiv.org/abs/2011.00637} {\path{arXiv:2011.00637}},
  \href {https://doi.org/10.21468/SciPostPhys.12.6.191}
  {\path{doi:10.21468/SciPostPhys.12.6.191}}.

\bibitem{He:2022xkh}
Song He, Zhang-Cheng Liu, and Yuan Sun.
\newblock {Entanglement entropy and modular Hamiltonian of free fermion with
  deformations on a torus}.
\newblock {\em JHEP}, 09:247, 2022.
\newblock \href {http://arxiv.org/abs/2207.06308} {\path{arXiv:2207.06308}},
  \href {https://doi.org/10.1007/JHEP09(2022)247}
  {\path{doi:10.1007/JHEP09(2022)247}}.

\bibitem{Cardona:2022cmh}
Biel Cardona and Javier Molina-Vilaplana.
\newblock {Entanglement renormalization of a $ \mathrm{T}\overline{\mathrm{T}}
  $-deformed CFT}.
\newblock {\em JHEP}, 07:092, 2022.
\newblock \href {http://arxiv.org/abs/2203.00319} {\path{arXiv:2203.00319}},
  \href {https://doi.org/10.1007/JHEP07(2022)092}
  {\path{doi:10.1007/JHEP07(2022)092}}.

\bibitem{Aramini:2022wbn}
Fabrizio Aramini, Nicol\`o Brizio, Stefano Negro, and Roberto Tateo.
\newblock {Deforming the ODE/IM correspondence with $
  \textrm{T}\overline{\textrm{T}} $}.
\newblock {\em JHEP}, 03:084, 2023.
\newblock \href {http://arxiv.org/abs/2212.13957} {\path{arXiv:2212.13957}},
  \href {https://doi.org/10.1007/JHEP03(2023)084}
  {\path{doi:10.1007/JHEP03(2023)084}}.

\bibitem{Hou:2023ytl}
Jue Hou, Miao He, and Yunfeng Jiang.
\newblock {$T\bar{T}$-deformed Entanglement Entropy for Integrable Quantum
  Field Theory}.
\newblock 6 2023.
\newblock \href {http://arxiv.org/abs/2306.07784} {\path{arXiv:2306.07784}}.

\bibitem{Jiang:2023ffu}
Xin Jiang, Peng Wang, Houwen Wu, and Haitang Yang.
\newblock {Timelike entanglement entropy and $T\bar{T}$ deformation}.
\newblock 2 2023.
\newblock \href {http://arxiv.org/abs/2302.13872} {\path{arXiv:2302.13872}}.

\bibitem{He:2023eap}
Song He, Jie Yang, Yu-Xuan Zhang, and Zi-Xuan Zhao.
\newblock {Pseudo-entropy for descendant operators in two-dimensional conformal
  field theories}.
\newblock 1 2023.
\newblock \href {http://arxiv.org/abs/2301.04891} {\path{arXiv:2301.04891}}.

\bibitem{He:2023wko}
Song He, Jie Yang, Yu-Xuan Zhang, and Zi-Xuan Zhao.
\newblock {Pseudo entropy of primary operators in
  $T\bar{T}$/$J\bar{T}$-deformed CFTs}.
\newblock 5 2023.
\newblock \href {http://arxiv.org/abs/2305.10984} {\path{arXiv:2305.10984}}.

\bibitem{Castro-Alvaredo:2023jbg}
Olalla~A. Castro-Alvaredo, Stefano Negro, and Fabio Sailis.
\newblock {Entanglement Entropy from Form Factors in
  $\mathrm{T}\bar{\mathrm{T}}$-Deformed Integrable Quantum Field Theories}.
\newblock 6 2023.
\newblock \href {http://arxiv.org/abs/2306.11064} {\path{arXiv:2306.11064}}.

\bibitem{Tian:2023fgf}
Jia Tian.
\newblock {On-shell action and Entanglement entropy of
  $\text{T}\bar{\text{T}}$-deformed Holographic CFTs}.
\newblock 6 2023.
\newblock \href {http://arxiv.org/abs/2306.01258} {\path{arXiv:2306.01258}}.

\bibitem{McGough:2016lol}
Lauren McGough, M\'ark Mezei, and Herman Verlinde.
\newblock {Moving the CFT into the bulk with $ T\overline{T} $}.
\newblock {\em JHEP}, 04:010, 2018.
\newblock \href {http://arxiv.org/abs/1611.03470} {\path{arXiv:1611.03470}},
  \href {https://doi.org/10.1007/JHEP04(2018)010}
  {\path{doi:10.1007/JHEP04(2018)010}}.

\bibitem{Araujo-Regado:2022gvw}
Goncalo Araujo-Regado, Rifath Khan, and Aron~C. Wall.
\newblock {Cauchy slice holography: a new AdS/CFT dictionary}.
\newblock {\em JHEP}, 03:026, 2023.
\newblock \href {http://arxiv.org/abs/2204.00591} {\path{arXiv:2204.00591}},
  \href {https://doi.org/10.1007/JHEP03(2023)026}
  {\path{doi:10.1007/JHEP03(2023)026}}.

\bibitem{Araujo-Regado:2022jpj}
Goncalo Araujo-Regado.
\newblock {Holographic Cosmology on Closed Slices in 2+1 Dimensions}.
\newblock 12 2022.
\newblock \href {http://arxiv.org/abs/2212.03219} {\path{arXiv:2212.03219}}.

\bibitem{Maldacena:2002vr}
Juan~Martin Maldacena.
\newblock {Non-Gaussian features of primordial fluctuations in single field
  inflationary models}.
\newblock {\em JHEP}, 05:013, 2003.
\newblock \href {http://arxiv.org/abs/astro-ph/0210603}
  {\path{arXiv:astro-ph/0210603}}, \href
  {https://doi.org/10.1088/1126-6708/2003/05/013}
  {\path{doi:10.1088/1126-6708/2003/05/013}}.

\bibitem{Nakata:2020luh}
Yoshifumi Nakata, Tadashi Takayanagi, Yusuke Taki, Kotaro Tamaoka, and Zixia
  Wei.
\newblock {New holographic generalization of entanglement entropy}.
\newblock {\em Phys. Rev. D}, 103(2):026005, 2021.
\newblock \href {http://arxiv.org/abs/2005.13801} {\path{arXiv:2005.13801}},
  \href {https://doi.org/10.1103/PhysRevD.103.026005}
  {\path{doi:10.1103/PhysRevD.103.026005}}.

\bibitem{Mollabashi:2020yie}
Ali Mollabashi, Noburo Shiba, Tadashi Takayanagi, Kotaro Tamaoka, and Zixia
  Wei.
\newblock {Pseudo Entropy in Free Quantum Field Theories}.
\newblock {\em Phys. Rev. Lett.}, 126(8):081601, 2021.
\newblock \href {http://arxiv.org/abs/2011.09648} {\path{arXiv:2011.09648}},
  \href {https://doi.org/10.1103/PhysRevLett.126.081601}
  {\path{doi:10.1103/PhysRevLett.126.081601}}.

\bibitem{Nishioka:2021cxe}
Tatsuma Nishioka, Tadashi Takayanagi, and Yusuke Taki.
\newblock {Topological pseudo entropy}.
\newblock {\em JHEP}, 09:015, 2021.
\newblock \href {http://arxiv.org/abs/2107.01797} {\path{arXiv:2107.01797}},
  \href {https://doi.org/10.1007/JHEP09(2021)015}
  {\path{doi:10.1007/JHEP09(2021)015}}.

\bibitem{Mollabashi:2021xsd}
Ali Mollabashi, Noburo Shiba, Tadashi Takayanagi, Kotaro Tamaoka, and Zixia
  Wei.
\newblock {Aspects of pseudoentropy in field theories}.
\newblock {\em Phys. Rev. Res.}, 3(3):033254, 2021.
\newblock \href {http://arxiv.org/abs/2106.03118} {\path{arXiv:2106.03118}},
  \href {https://doi.org/10.1103/PhysRevResearch.3.033254}
  {\path{doi:10.1103/PhysRevResearch.3.033254}}.

\bibitem{Narayan:2022afv}
K.~Narayan.
\newblock {de Sitter space, extremal surfaces and ''time-entanglement''}.
\newblock 10 2022.
\newblock \href {http://arxiv.org/abs/2210.12963} {\path{arXiv:2210.12963}}.

\bibitem{Doi:2022iyj}
Kazuki Doi, Jonathan Harper, Ali Mollabashi, Tadashi Takayanagi, and Yusuke
  Taki.
\newblock {Pseudoentropy in dS/CFT and Timelike Entanglement Entropy}.
\newblock {\em Phys. Rev. Lett.}, 130(3):031601, 2023.
\newblock \href {http://arxiv.org/abs/2210.09457} {\path{arXiv:2210.09457}},
  \href {https://doi.org/10.1103/PhysRevLett.130.031601}
  {\path{doi:10.1103/PhysRevLett.130.031601}}.

\bibitem{Doi:2023zaf}
Kazuki Doi, Jonathan Harper, Ali Mollabashi, Tadashi Takayanagi, and Yusuke
  Taki.
\newblock {Timelike entanglement entropy}.
\newblock 2 2023.
\newblock \href {http://arxiv.org/abs/2302.11695} {\path{arXiv:2302.11695}}.

\bibitem{Narayan:2023ebn}
K.~Narayan and Hitesh~K. Saini.
\newblock {Notes on time entanglement and pseudo-entropy}.
\newblock 3 2023.
\newblock \href {http://arxiv.org/abs/2303.01307} {\path{arXiv:2303.01307}}.

\bibitem{Kawamoto:2023nki}
Taishi Kawamoto, Shan-Ming Ruan, Yu-ki Suzuki, and Tadashi Takayanagi.
\newblock {A Half de Sitter Holography}.
\newblock 6 2023.
\newblock \href {http://arxiv.org/abs/2306.07575} {\path{arXiv:2306.07575}}.

\bibitem{Chu:2023zah}
Chong-Sun Chu and Himanshu Parihar.
\newblock {Time-like entanglement entropy in AdS/BCFT}.
\newblock {\em JHEP}, 06:173, 2023.
\newblock \href {http://arxiv.org/abs/2304.10907} {\path{arXiv:2304.10907}},
  \href {https://doi.org/10.1007/JHEP06(2023)173}
  {\path{doi:10.1007/JHEP06(2023)173}}.

\bibitem{Ryu:2006bv}
Shinsei Ryu and Tadashi Takayanagi.
\newblock {Holographic derivation of entanglement entropy from AdS/CFT}.
\newblock {\em Phys. Rev. Lett.}, 96:181602, 2006.
\newblock \href {http://arxiv.org/abs/hep-th/0603001}
  {\path{arXiv:hep-th/0603001}}, \href
  {https://doi.org/10.1103/PhysRevLett.96.181602}
  {\path{doi:10.1103/PhysRevLett.96.181602}}.

\bibitem{Ryu:2006ef}
Shinsei Ryu and Tadashi Takayanagi.
\newblock Aspects of holographic entanglement entropy.
\newblock {\em Journal of High Energy Physics}, 2006(08):045--045, aug 2006.
\newblock \href {https://doi.org/10.1088/1126-6708/2006/08/045}
  {\path{doi:10.1088/1126-6708/2006/08/045}}.

\bibitem{Witten:2022xxp}
Edward Witten.
\newblock {A Note On The Canonical Formalism for Gravity}.
\newblock 12 2022.
\newblock \href {http://arxiv.org/abs/2212.08270} {\path{arXiv:2212.08270}}.

\bibitem{Brown1986}
J.~David Brown and M.~Henneaux.
\newblock Central charges in the canonical realization of asymptotic
  symmetries: An example from three-dimensional gravity.
\newblock {\em Commun. Math. Phys.}, 104:207--226, 1986.
\newblock \href {https://doi.org/10.1007/BF01211590}
  {\path{doi:10.1007/BF01211590}}.

\bibitem{Calabrese:2004eu}
Pasquale Calabrese and John~L. Cardy.
\newblock {Entanglement entropy and quantum field theory}.
\newblock {\em J. Stat. Mech.}, 0406:P06002, 2004.
\newblock \href {http://arxiv.org/abs/hep-th/0405152}
  {\path{arXiv:hep-th/0405152}}, \href
  {https://doi.org/10.1088/1742-5468/2004/06/P06002}
  {\path{doi:10.1088/1742-5468/2004/06/P06002}}.

\bibitem{Calabrese:2009qy}
Pasquale Calabrese and John Cardy.
\newblock {Entanglement entropy and conformal field theory}.
\newblock {\em J. Phys. A}, 42:504005, 2009.
\newblock \href {http://arxiv.org/abs/0905.4013} {\path{arXiv:0905.4013}},
  \href {https://doi.org/10.1088/1751-8113/42/50/504005}
  {\path{doi:10.1088/1751-8113/42/50/504005}}.

\bibitem{Narayan:2015vda}
K.~Narayan.
\newblock {Extremal surfaces in de Sitter spacetime}.
\newblock {\em Phys. Rev. D}, 91(12):126011, 2015.
\newblock \href {http://arxiv.org/abs/1501.03019} {\path{arXiv:1501.03019}},
  \href {https://doi.org/10.1103/PhysRevD.91.126011}
  {\path{doi:10.1103/PhysRevD.91.126011}}.

\bibitem{Narayan:2015oka}
K.~Narayan.
\newblock {de Sitter space and extremal surfaces for spheres}.
\newblock {\em Phys. Lett. B}, 753:308--314, 2016.
\newblock \href {http://arxiv.org/abs/1504.07430} {\path{arXiv:1504.07430}},
  \href {https://doi.org/10.1016/j.physletb.2015.12.019}
  {\path{doi:10.1016/j.physletb.2015.12.019}}.

\bibitem{Narayan:2016xwq}
K.~Narayan.
\newblock {On $dS_4$ extremal surfaces and entanglement entropy in some ghost
  CFTs}.
\newblock {\em Phys. Rev. D}, 94(4):046001, 2016.
\newblock \href {http://arxiv.org/abs/1602.06505} {\path{arXiv:1602.06505}},
  \href {https://doi.org/10.1103/PhysRevD.94.046001}
  {\path{doi:10.1103/PhysRevD.94.046001}}.

\end{thebibliography}

\end{document}